\documentclass[10pt]{article}
\usepackage{graphicx}
\begin{document}

\title{The investigation of the Orientation of Galaxies in Structures}

\medskip \author {Elena Panko ${^1}$, Paulina Piwowarska ${^2}$, Jolanta God{\l}owska$^{3}$,\\
 W{\l}odzimierz God{\l}owski${^4}$, Piotr Flin $^{5}$}

\maketitle
 
1. Kalinenkov Astronomical Observatory, Nikolaev National University,
Nikolskaya, 24,  54030 Nikolaev, Ukraine  email: panko.elena@gmail.com 

2. Uniwersytet Opolski, Institute of Physics, ul.  Oleska  48,
45-052 Opole, Poland e-mail: paoletta@interia.pl
 
3. Department of Monitoring and Modelling Air Pollution, Institute of Meteorology
and Water Management, 30-215 Krakow,  Borowego 14, Poland  email: Jolanta.Godlowska@imgw.pl

4. Uniwersytet Opolski, Institute of Physics, ul.  Oleska  48,
45-052 Opole, Poland e-mail: godlowski@uni.opole.pl

5. Institute of Physics, Jan Kochanowski University, 25-406 Kielce,
 Swietokrzyska 15, Poland email: sfflin@cyf-kr.edu.pl

\section*{Abstract}
\medskip

The investigation of the orientation of galaxies is a standard test concerning to scenarios of 
galaxy formation, because different theories of galaxy formation make various 
predictions regarding to the angular momentum of galaxies. The new method of analysis of 
the alignment of galaxies in clusters was proposed in the paper \cite{g12} and now 
is improved. We analyzed the distribution of the position angles of the galaxy major axes,
as well as the distribution of two angles describing the spatial orientation of galaxy plane,
which gives the information about galaxy angular momenta. We discuss the orientation of galaxies 
in groups and clusters of galaxies. The results show the dependence of the alignment with respect 
to clusters richness. The implications of the results for theories of galaxy formation are 
discussed as well.

{\bf keywords}
 angular momenta, galaxies

\section{Introduction}

Different theories of galaxy formation (for example 
\cite{Peebles69,Zeldovich70,Sunyaew72,Doroshkevich73,Shandarin74,Dekel85,Wesson82,Silk83,Bower06})
 make  different predictions regarding to the angular momentum of galaxies. These classical theories 
of formation of galaxies and its structures were then revised and improved by many researchers (for example 
\cite{Lee00,Lee01,Lee02,Navarro04,Mo05,Brook08,VC11,Paz11,Shandarin12,Giahi13,t06,Codis12,Giahi13}) 
but this statement remains valid. More generally, the observed variations in angular momentum represent 
simple but fundamental constraints for any model of galaxy formation \cite{Rom12}.

In the commonly accepted $\Lambda$CDM model, the Universe deems to be spatially
flat, as well as homogeneous and isotropic at the same appropriate scale.  In this model 
the structures were formed from the primordial adiabatic, nearly scale invariant 
Gaussian random fluctuations (\cite{Silk68,Peebles70,Sunyaew70}). This picture 
is in agreement with both the numerous numerical simulations \cite{Springel05,w1,w2,Codis12}
and the observations. Unfortunately the angular momenta of galaxies are known only for very 
few galaxies. Therefore instead of the angular momenta, the orientation of galaxies 
is investigated. In order to this either the orientation of position angles of major
galaxy axis \cite{h4}, or spatial orientation of galaxy planes \cite{Oepik70,Jaaniste78,f4} 
is investigated.

\cite{g05} suggest that the alignment should increase with richness
of the cluster. 

It also should be noticed that it is commonly agree that for groups and clusters of galaxies
 there is no evidence of rotation  and groups and clusters of galaxies do not rotate (for example 
\cite{Regos89,Diaferio97,Diaferio99,Rines03} - see however \cite{Kal05} for opposite opinion).  
Moreover recently \cite{Hwang07} examined dispersions and  velocity gradient of 899 Abell clusters 
and found a possible rotation in only six of them. Thus, any non-zero angular momentum of groups and 
clusters of galaxies would just come from possible alignment of galaxy spins.

The idea that richer clusters we can expect more strong alignments is results both from  analysis of
 implications of theoretical relations between the angular momentum and the mass of the structure and 
from analysis of the observational results of alignment of galaxies in different scales. It is of course
 true that presently it is commonly believed that galaxies are formed before clusters but our idea is not
 in conflict with this. It can be explained both in the tidial-torque scenario \cite{HP88,Catelan96,Fedeli13} 
and in the Li model \cite{Li98}.

The prediction that the alignment should increase with richness of the cluster was  confirmed  by  
\cite{Aryal07} but both \cite{g05} and \cite{Aryal07} analysis was qualitative only.
Therefore this problem was analyzed in details in later papers
\cite{g10a,g11a,g11b,g12}. In this works it was  found that  degree of the 
alignment of galaxies orientations in  clusters depend on theirs number of members 
and increase with the amount of the galaxies' members. It is equivalent to the existence 
of a relation between anisotropy and the number of galaxies in a cluster. Moreover it was 
found that orientations of galaxies analyzed sample of rich Abell galaxy clusters are not 
random, i.e., that there exists an alignment of galaxies in rich Abell galaxy clusters. 
In the present paper we would like to investigated this problem deeper, improving new method
of analysis of the alignment of galaxies in clusters proposed in the paper \cite{g12}.

\begin{table}[h]
\begin{center}
{\scriptsize
\caption {The results of numerical simulations for positions angles $P$, Tullys' groups.}
\label{tab:t1}
\begin{tabular}{|c|c|c|c|c|}
\hline
Test&$\bar{x}$&$\sigma(x)$&$\sigma(\bar{x})$&$\sigma(\sigma(x))$\\
\hline
  $\chi^2$                         & 16.9524&  1.4592&   0.0461&  0.0326 \\
  $\Delta_{1}/\sigma(\Delta_{1})$  &  1.2513&  0.1543&   0.0048&  0.0034 \\
  $\Delta/\sigma(\Delta)$          &  1.8772&  0.1581&   0.0050&  0.0035 \\
  $C$                              & -1.0256&  0.9295&   0.0294&  0.0208 \\
  $\lambda$                        &  0.7317&  0.0615&   0.0019&  0.0014 \\
  $\Delta_{11}/\sigma(\Delta_{11})$&  0.0065&  0.2346&   0.0074&  0.0052 \\
\hline
\end{tabular}
}
\end{center}
\end{table}

\begin{table}[h]
\begin{center}
{\scriptsize
\caption {The statistics of the observed distributions for real Tullys' groups.}
\label{tab:t2}
\begin{tabular}{|c|cc|cc|cc|}
\hline
\multicolumn{1}{|c}{}&
\multicolumn{2}{|c}{$P$}&
\multicolumn{2}{|c}{$\delta_D$}&
\multicolumn{2}{|c|}{$\eta$}\\
\hline
Test&$\bar{x}$&$\sigma(x)$&$\bar{x}$&$\sigma(x)$&$\bar{x}$&$\sigma(x)$\\
\hline
$\chi^2$                         & 16.800& 1.152& 17.983& 1.601& 17.267& 1.579\\
$\Delta_{1}/\sigma(\Delta_{1})$  &  1.218& 0.176&  1.356& 0.116&  1.509& 0.208\\
$\Delta/\sigma(\Delta)$          &  2.081& 0.218&  2.040& 0.172&  2.343& 0.205\\
$C$                              &  0.026& 0.932&  0.044& 0.992&  1.268& 0.984\\
$\lambda$                        &  0.834& 0.072&  0.740& 0.055&  1.829& 0.064\\
$\Delta_{11}/\sigma(\Delta_{11})$& -0.283& 0.182&  0.292& 0.237&  0.773& 0.267\\
\hline
\end{tabular}
}
\end{center}
\end{table}

\section{Overview}

The orientation of galaxies is usually investigated either by analysing the 
distribution of the position angle of the galactic major axies or by the 
analysis of the spatial orientation of the normal line to the galaxy main 
plane in the investigated coordinate system. To obtain the spatial orientation
of galaxies for each galaxy, two angles are
determined: $\delta_D$ - the angle between the normal to the galaxy plane
and the main plane of the coordinate system, and $\eta$ - the angle between
the projection of this normal onto the main plane and the direction towards
to the zero initial meridian. 

In our previous papers we've shown that analysis of the spatial orientation of
galaxies' planes can be used as a general, standard test of galaxies forming 
scenario \cite{f4,g2,g3,g5,g10a,f11,g11a}. Any study of galactic orientation 
based on the projection of galaxies on the celestial sphere gives
a four-hold ambiguity in the solution for angular momentum. By the reason of
none information connected with the direction of the galaxy spin our analysis
is reduced to only two solutions. Using the Supergalactic coordinate system
(\cite{f4} based on \cite{TS76}) the following relations
between angles ($L$, $B$, $P$) and ($\delta_D$, $\eta$) hold:
\begin{equation}
\sin\delta_D  =  -\cos{i}\sin{B} \pm \sin{i}\cos{r}\cos{B},
\end{equation}
\begin{equation}
\sin\eta  =  (\cos\delta_D)^{-1}[-\cos{i}\cos{B}\sin{L} + \sin{i}
(\mp \cos{r}\sin{B}\sin{L} \pm \sin{r}\cos{L})],
\end{equation}
\begin{equation}
\cos\eta  =  (\cos\delta_D)^{-1}[-\cos{i}\cos{B}\cos{L} + \sin{i}
(\mp \cos{r}\sin{B}\cos{L} \mp \sin{r}\sin{L})],
\end{equation}
where $r=P-\pi/2$.
As a result of the reduction of our analysis into two solutions it is necessary
to consider the sign of the expression:
$S=-\cos{i}\cos{B} \mp \sin{i}\cos{r}\sin{B}$
and for $S\ge 0$ reverse sign of $\delta_D$ respectively\cite{g10a}.

\begin{table}
  \begin{center}
\caption{The results of numerical simulation - sample of 247 cluster
each with number of members galaxies the same as in the real cluster.}
  \label{tab:t3}
 {\scriptsize
  \begin{tabular}{|l|cc|cc|cc|}
  \hline 
\multicolumn{1}{|c}{}&
\multicolumn{2}{|c}{$P$}&
\multicolumn{2}{|c}{$\delta_D$}&
\multicolumn{2}{|c|}{$\eta$}\\
\hline
Test&$\bar{x}$&$\sigma(x)$&$\bar{x}$&$\sigma(x)$&$\bar{x}$&$\sigma(x)$\\
\hline
$\chi^2$                         &$34.9798$&$0.5364$&$35.5824$&$0.5461$&$36.3663$&$0.5332$\\
$\Delta_{1}/\sigma(\Delta_{1})$  &$ 1.2550$&$0.0419$&$ 1.2523$&$0.0428$&$ 1.3756$&$0.0473$\\
$\Delta/\sigma(\Delta)$          &$ 1.8788$&$0.0436$&$ 1.8844$&$0.0492$&$ 2.0208$&$0.0518$\\
$C$                              &$-1.0195$&$0.3749$&$-0.5226$&$0.3807$&$-0.1667$&$0.4043$\\
$\lambda$                        &$ 0.7720$&$0.0168$&$ 0.8099$&$0.0201$&$ 0.8193$&$0.0188$\\
$\Delta_{11}/\sigma(\Delta_{11})$&$ 0.0014$&$0.0645$&$ 0.0039$&$0.0151$&$ 0.0083$&$0.0705$\\
\hline
  \end{tabular}
  }
 \end{center}
\vspace{1mm}
\end{table}

\begin{figure}[b]
\begin{center}
 \includegraphics[angle=270,scale=0.23]{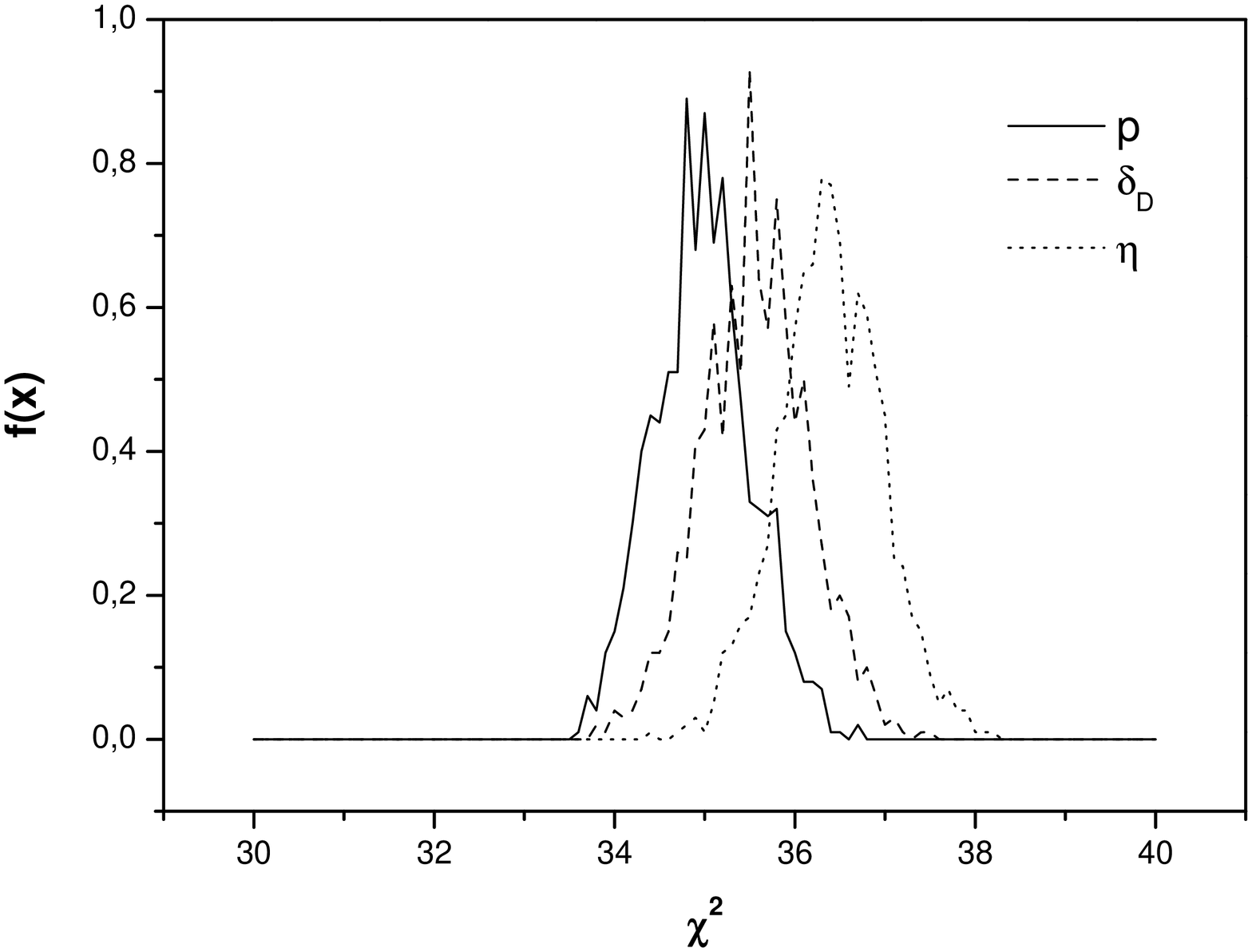}
 \includegraphics[angle=270,scale=0.23]{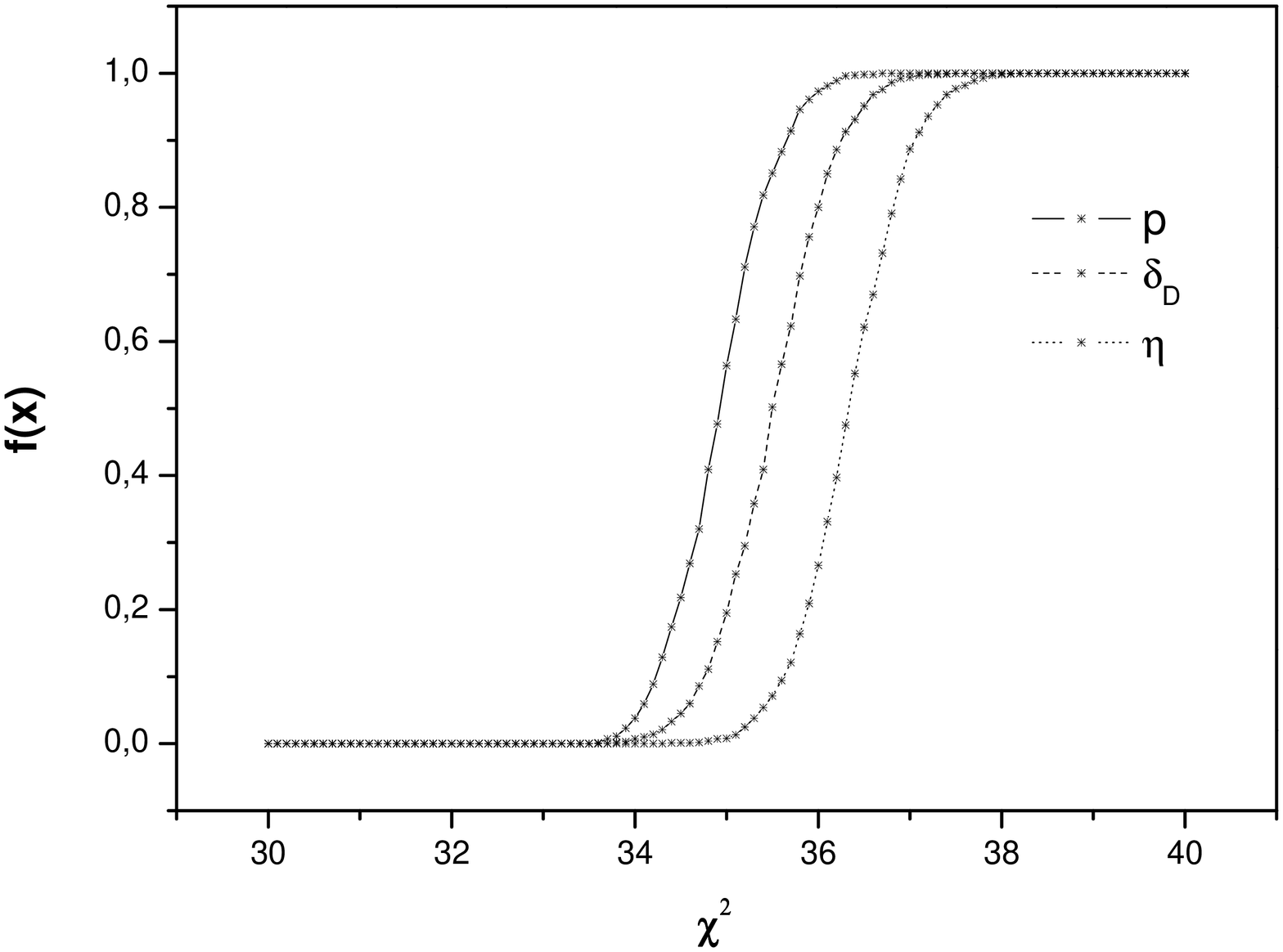}
 \caption{Differences in Probability Distribution Function (PDF) (left panel) and
 Cumulative Distribution Function (CDF) (right panel),
  for angles $P$, $\delta_d$ and $\eta$ ($\chi^2$ statistics).
The figure was obtained from 1000 simulations of sample of 247 clusters
each with number of member galaxies the same as in the real cluster.}
   \label{fig1}
\end{center}
\end{figure}

\begin{table}
\begin{center}
\caption {The statistics of the distributions for position angles for sample of 247 Abell clusters.}
\label{tab:t4}
{\scriptsize
\begin{tabular}{|c|cc|cc|}
\hline
\multicolumn{1}{c}{}&
\multicolumn{2}{c}{Equatorial coordinates}&
\multicolumn{2}{c}{Supergalactic coordinates}\\
\hline
Test&$\bar{x}$&$\sigma(\bar{x})$&$\bar{x}$&$\sigma(\bar{x})$\\
\hline
$\chi^2$                         &$36.8591$&$0.5924$&$36.7899$&$0.6315$\\
$\Delta_{1}/\sigma(\Delta_{1})$  &$ 1.7046$&$0.0622$&$ 1.7021$&$0.0626$\\
$\Delta/\sigma(\Delta)$          &$ 2.2663$&$0.0594$&$ 2.2746$&$0.0591$\\
$C$                              &$ 1.1940$&$0.4530$&$ 1.1220$&$0.4237$\\
$\lambda$                        &$ 0.9177$&$0.0240$&$ 0.9138$&$0.0220$\\
$\Delta_{11}/\sigma(\Delta_{11})$&$-0.0005$&$0.0855$&$ 0.0940$&$0.0924$\\
\hline
\end{tabular}
}
\end{center}
\end{table}

Significant progress in the investigation of galaxy plane orientation was made by \cite{h4}. 
They discussed in detailed manner the method of investigating the galaxies' orientation
through analysing distribution of position angles as well as the influence of possible errors 
and observational effects. \cite{h4} analysed the distributions of position angles using the 
$\chi^2$-test, Fourier test and the autocorrelation test. Since \cite{h4} this method was 
accepted as standard method for invetigation of a galactic alignment.
Thera are several modification of the original \cite{h4} method \cite{f4,Kindl87,g2,g3,g5,Ar00,g10a,g12}
 
In order to detect non-random effects in the distribution of the
investigated angles: $\delta_D$, $\eta$ and $P$ we divided the entire
range of the analyzed angles into $n$ bins and carried
out three different statistical tests. These tests were: the $\chi^2$ test,
the autocorrelation test and the Fourier test  \cite{h4,g10a,g11b,g12}.
We analyzed two  samples of galaxy clusters. We computed the value of analyzed statistics
for each cluster and later the mean value of the analyzed statistics. 
It was compared with theoretical predictions and results of numerical simulations obtained with using 
RANLUX generator \cite{Luescher94,Luescher10}. We used 6 statistics discussed in detail in \cite{g12}.

At first we analyzed the sample of 18 Tullys' clusters (see also \cite{g11b}). The results are 
presented in  the Tables 1 and 2. One should note, that there are insignificant differences
with results in Table 2 and results in \cite{g11b} (Table 5) because of a small differences
in method of computing the inclination angle with taking into account "true" axial ratio
$q_0$ which depends on morphological type and converting axial ratio $q$ to standard photometrical 
axial ratio (see \cite{fp85}). From our results it is clearly seen that for samples of 18 groups 
of galaxies we do not found any significant alignments. 

The next step was the analysis of the distribution of the position angles in the sample 
of  rich Abell clusters \cite{g10a,g12}.  This sample was taken from \cite{Panko06} catalog,
which is catalog of galaxy cluster obtain on the  base of the Muenster Red Sky Survey  \cite{MRSS03}.
Our sample containss 247 Abell clusters with richness at least 100 galaxies each being 
identified with one of ACO clusters \cite{ACO}. The results of our investigation are presented 
in the Tables 3 and 4. Now, on the contrary to the previous investigated samples of Tully groups, 
we found significant deviation of mean values of the statistics from expected values obtained from 
numerical simulations. From the Figure 1 and Table 3 we could find that results of the expected value 
of analyzed statistics for angles $\delta_D$ and $\eta$ are larger than that obtained for position 
angles $p$. It is mostly caused of the fact that during the process of deprojection of the spatial
orientation of galaxies from its optical images we obtain two possible orientations - see equations 1-3.
From analysis of these equations it is easy to see that both solutions are not independent and as a  
result the distribution of analyzed statistics is modified and must be obtained from numerical 
simulations. Such differences have not changed our conclusion obtained from analysis of Tullys' groups 
of galaxies, because we have not found significant deviation even if we took smaller values obtained 
from simulations of position angles, but it should be taken into account during future analysis
of the spatial orientation of galaxies in our sample of rich Abell clusters.

\section{Conclusions}

In the present paper we confirmed the predictions from the previous analysis \cite{g10a,g11b,g12} 
which showed the dependency of the alignment of galaxies in clusters on richness of the 
cluster. Moreover, we found that for the sample of Tullys' groups of galaxies we do not found 
any significant alignment while it is observed for the sample of 247 Abells' clusters, with 
richness at least 100 galaxies each. In the present paper  we extended our analysis by
showing that expected values of analyzed statistics for angles giving spatial 
orientation of galaxies, $\delta_D$ and $\eta$ are greater than that obtained for 
position angles of major axis of galaxies. It is mostly because during  the analysis of 
the spatial orientation of galaxies we have two possible orientations of galaxies which 
are not independent to each other. Our results lead to the conclusion  that the angular 
momentum of the cluster increases with the mass of the structure. With such a dependency 
it is natural to expect that in rich clusters significant alignment should be present.  
Usually a dependency between the angular momentum and the mass of the
structure is presented as empirical relation $J\sim M^{5/3}$
\cite{Wesson79,Carrasco82,White84,Brosche86,Paz08,Rom12}, for rewiev see also \cite{Sch09}. 
In our opinion the observed
relation between the richness of the galaxy cluster and the alignment is due
to tidal torque, as suggested by \cite{HP88} and \cite{Catelan96}. Moreover, the
analysis of the linear tidal torque theory is pointing in the same direction
\cite{Noh06a,Noh06b}. They've noticed the connection of the alignment with the
considered scale of the structure.  It should be also noticed that many of other result 
conected with galaxy aligmenth is interpreted in the light of tidial torque theory.
For example  Gonzalez and Teodoro \cite{Gon10}  interpreted the 
alignment of just the brightest galaxies within a cluster as an effect of action 
of gravitational tidal forces. However one should note that our result, increase of 
aligmenth with cluster richness is also compatible with the prediction of the Li model 
in which galaxies are formed in the rotating universe (\cite{Li98,g03,g05}).

We should keep in mind that Li's model \cite{Li98} remains valid only provided rotation 
occurs on a sufficiently large scale. Thus in this model considerations concerning 
angular momenta of galaxy structures will be also valid in the case of large scale, 
but not necessarily global, rotation of the Universe. It is important because observed 
the amplitude of quadrupole fluctuations value provided by both COBE and WMAP measurements 
is only $\Delta T^2=249 \mu K^2$ \cite{Hinshaw07}. More generally the large scale anisotropy linked with 
the rotation of the Universe, homogeneous magnetic field, anisotropy of curvature and 
other similar factors is strongly restricted by WMAP observations of quadrupole anisotropy 
of relic radiation (see for review Demianski \& Doroshkevich \cite{DD07}.One should note however 
that most serious problem of Li's model \cite{Li98} is consists in the fact that the observed 
amount of rotation of spiral galaxies cannot arise from the Universe's rotation alone, 
since the required amount of rotation of the Universe on the order obtained earlier by 
Birch \cite{Birch82,Birch83} which is too large in comparison with the detected anisotropies of cosmic 
background radiation (for review see \cite{g11a}

Recently,  there  have  been  also some  attempts to  investigate galaxy  angular 
momenta  on a large scale.  \cite{Paz08} analysing galaxies from the  Sloan  Digital 
Sky Survey  catalogue  found  that the  galaxy  angular  momenta  are aligned  perpendicularly 
to the planes of large-scale structures, while  there is no such effect for the low-mass 
structures. They interpret this as consistent with their simulations based on  the
mechanism  of  tidal interactions. The change of alignment with the surrounding neigbourhood
was observed also in alignment study in void vicinity \cite{v11} being continuation of earlier 
study of galaxy orientation in regions surrounding bubble-like voids \cite{t06}.
\cite{Jones10}  found that the  spins  of  spiral  galaxies located 
within cosmic web filaments tend to be aligned along  the larger  axis of the filament, 
which they interpreted as  "fossil" evidence indicating that the action of large 
scale tidal  torques effected the alignments of galaxies located in cosmic filaments.
\cite{Tempel13} found evidence that the spin axes of bright spiral galaxies have a weak tendency
to be aligned parallel to filaments. For elliptical/S0 galaxies, they  have a statistically
significant result that the spin axes of ellipticals are aligned preferentially perpendic-
ular to the host filaments. Lee \cite{Lee11} comparing of his observational results with the 
analytic model based on the tidal torque theory reveals that the spin correlation function 
for the late-type spiral galaxies follow the quadratic scaling of the linear density correlation 
and that the intrinsic correlations of the galaxy spin axes are stronger than that of the underlying 
dark halos. The intrinsic correlations between galaxy spins and intermediate principal axes of the 
tidal shears was also found by Lee and Erdogdu \cite{Lee07}.
 
Possible relation between filament and orientation of galaxies was noticed also by  
God³owski \& Flin \cite{g10}. In this paper the orientation of galaxy groups in the Local Supercluster 
was studied, and it was found a strong alignment of the major axis of the groups with directions 
towards the supercluster center (Virgo cluster) as well as with the line joining the two brightest 
galaxies in the group. The Interpretation of these observational results was following. The brightest 
galaxies of the group, believed to be the most massive ones, originated first. Afterward, the hierarchical 
clustering leads to aggregation of galaxies around these two galaxies. The groups are formed on the same or 
similarly oriented filaments. It should be also noticed that galaxy cluster intrinsic alignments to very 
large scales of $100h^{-1}$Mpc, representing a tendency of clusters to point preferentially towards other 
clusters was found by Smargon et al.\cite{Smargon12} .

Possible significance of the evolution of the alignmenta with 
redshift is sugest by the results of the paper of \cite{Song12}, which found that the alignment 
profile of cluster galaxies drops faster at higher redshifts and on smaller mass scales.
Moreover, one should note that the largest scale alignment was found in the series of paper 
by Hutsemekers (\cite{Hutsemekers98,Hutsemekers01,Hutsemekers05}) during analyzis of the  
alignment of quasar polarization vectors.The polarization vectors appear to be coherently oriented 
or aligned over huge (about 1 Gpc) regions at the sky. Furthermore, the mean polarization angle 
$\theta$ appears to rotate with redshift at the rate of about 300 per Gpc. These results usually 
are not questioned, (with exception of Joshi et al. \cite{Joshi07} which not found any effects during 
analysis of theirs sample \cite{Jackson07}), however the origin of this effect is still discussed. 
While interpretations like a global rotation of the Universe 
can potentially explain the effect the properties they observed qualitatively correspond to the dichroism 
and birefringence predicted by the photon-pseudoscalar oscillation within a magnetic field. The possible 
interpretations was discussed many times for example by Hutsemekers 
\cite{Hutsemekers01,Hutsemekers05,Hutsemekers10} and recently by Agarwal,Kamal and Jain \cite{Agarwal11}

In our further paper we would like to extend our consideration to more detailed analysis
of the distribution of two angles $\delta_D$ and $\eta$, describing the spatial orientation of
the galaxy plane. Moreover, we would like to investigate does the effect found in the present 
paper depends on the cluster BM type and velocity dispersion of member galaxies.

\section*{Acknowledgments}
 
This research has made use of the NASA/IPAC Extragalactic Database (NED)
which is operated by the Jet Propulsion Laboratory, California Institute
of Technology, under contract with the National Aeronautics and Space
Administration.

\end{document}